# Enhanced thermoelectric performance in Ca substituted $Sr_3SnO$

Enamul Haque and M. Anwar Hossain*

Department of Physics, Mawlana Bhashani Science and Technology University,

Santosh, Tangail-1902, Bangladesh

Email: anwar647@mbstu.ac.bd

**Abstract**

We report 45% enhancement in the thermoelectric figure of merit, ZT of $Sr_3SnO$ via Ca substitution. First-principles calculations have been performed to study the electronic and thermoelectric transport properties in Ca substituted $Sr_3SnO$ ($Sr_{3-x}Ca_xSnO$, $0 \leq x \geq 3$). The effects of Ca subtitution on bandgap are studied and detailed mechanisms are proposed to explain the obtained results. We have found that effective mass and thermopower of $Sr_3SnO$ redueces with the increase of hole concentration. The optimum hole concentration has been obtained for $Sr_2CaSnO$ and the corresponding Seebeck coefficient is 219 µV/K. The electrical conductivity of $Sr_3SnO$ and its alloys exhibits semiconducting nature which contradicts with experimental results in $Ca_3SnO$. We have found that due to the Ca-deficiency, the $Ca_3SnO$ shows the metallic conductivity and removes this contradiction with our results. The lattice thermal conductivities ($\kappa_l$) of $Sr_3SnO$ and $Ca_3SnO$ have been calculated by using both PBE and GW functionals. The lattice thermal conductivity obtained by PBE functional largely underestimates the experimental value for $Ca_3SnO$. The total thermal conductivity (with $\kappa_l$ obtained by GW) at 300K is 2.33 and 1.897 W/mK for $Sr_3SnO$ and $Ca_3SnO$, respectively, with excellent agreement with experimental value 1.707 W/mK for $Ca_3SnO$. The dimensionless figure of merit (ZT) for $Sr_2CaSnO$ at 500 K is 0.6 and making it promising for thermoelectric applications.

1. Introduction

The antiperovskite Dirac metal oxides (ADMOs) have attracted much attention of researchers due to their unusual superconductivity [1–3], topological behavior [4], ferromagnetic [5–9] and good thermoelectric transport properties [10,11]. In these oxides, the metal ions reverse their positions to oxygen ions and thus metal ions have negative valence states [12]. The $Sr_3SnO$ (SSO), an ADMO, has been experimentally found to exhibit topological superconductivity with asuperconducting transition temperature ($T_c$) ~5 K due to Sr deficiency[1,3]. The strong hybridyzation between Sr-4d and Sn-5p orbitals causes topological superconductivity to be present in SSO[1,2]. However, many theoretical studies reported that the ADMOs exhibit topological insulating and semiconducting behavior due to the band inverion between two quartets[4]. The SSO with Si (001) has found to be dilute magnetic semiconductor [5,7]. The most recent theoretical studies on $A_3SnO$ ($A$ = Ca, Sr, Ba) have been found that these ADMOs exhibit good thermoelectric properties and are suitable for high temperature thermoelectric devices[11]. The $Ca_3SnO$ (CSO) has been experimentally found to possesses high Seebeck coefficient, approaching about ~ 100 µV/K at the room temperature[10]. Since the co-doping can significantly improve the thermoelectric performance by reducing thermal conductivity and enhancing Seebeck coefficient [12–15], it is reasonable to study of the thermoelectric properties in Ca substituted $Sr_3SnO$. The thermoelectric device requires high-performance thermoelectric materials to convert heat energy into electricity efficiently. The performance of thermoelectric materials at a given temperature T can be calculated by using the equations [16], $ZT = \frac{S^2 \sigma T}{\kappa_e + \kappa_l}$, where $S$, $\sigma$, $\kappa_e$, $\kappa_l$ are the Seebeck coefficient, eletrical conductivity, eletronic and lattice thermal conductivity. Thus, ZT will be high for a material possessing high thermopower and low thermal conductivity. However, it is very difficult to opimize these parameters simultaneously. The recent experimental study on a polycrystalline sample of $Ca_3SnO$ has been revealed that the thermal conductivity of $Ca_3SnO$ is relatively low (1.7 W/mK at 290 K) and $Ca_3SnO$ shows metallic conductivity but experimentlists did not mention why the sample shows metallic conductivity [10]. Many theoretical studies on this compounds have been revealed that $Ca_3SnO$ is a topological insulator [4,11]. Therefore, it is important to study the electrical conductivity of $Ca_3SnO$ to find the reason behind this contradictory between experimental and theoretical results. The SSO is a cubic crystal with experimental lattice paramter a=5.1394 Å and space group $Pm\bar{3}m$ (#221). The Wyckoff positions for Sr, Sn and O atoms are 3c (0, 0.5, 0.5), 1a (0, 0, 0) and 1b (0.5, 0.5, 0.5), respectively [17,18].

In this paper, we report the electronic and thermoelectric transport properties of Ca substituted $Sr_3SnO$ by using the first-principles calculations. We also study the role of Ca deficiency on the thermoelectric transport properties of $Ca_3SnO$. We find that the thermoelectric performance is significantly improved by Ca-doped in SSO. We also find that the $Ca_3SnO$ shows the metallic conductivity for small Ca-deficiency as experimentally observed [10]. Moreover, the GW method provides reasonably accurate lattice thermal conductivity than PBE-GGA method.

2. **Computaional details**

Electronic structure was studied by using a full potential augmented linearized plane wave (FP-LAPW) method in WIEN2k [19]. The structural optimization was performed by using the experimental lattice parameters 5.1394Å [18] and the muffin tin radii: 2.3, 2.5, 2.3, and 2.2 Bohr for Sr, Sn, O and Ca, respectively. For good convergence of basis set, the kinetic energy cutoff ($RK_{max}$) was set to 7.0. To overcome the underestimation of the bandgap (PBE -GGA [20]), Tran-Blaha modified Becke-Johnson potential (TB-mBJ) [21,22] was used. For self-consistent field (SCF) and density of states as well as bandstructure calculaion, $15 \times 15 \times 15$ and $21 \times 21 \times 21$ k-point were used, respectivley. To generate the required data for the transport properties calculation, SCF calculation was performed again by using a finer $47 \times 47 \times 47$ k-point. The transport properties were calculated by solving semi-classsical Boltzmann transport equation as implemented in BoltzTraP [23]. The BoltzTraP calculates the electron transport parameters within the constant relaxation time approximation. The values of transport parameters were taken at the Fermi level of 0K. The lattice thermal conductivity was calculated by solving linerized phonon Boltzmann equation (LBTE) withinthe finite displacement approach [24] as implemented in Phono3py [25]. In this code, the single-mode relaxation time (SMRT) approxationis is used. This calculation was performed by creating $2 \times 2 \times 2$ supercell that contains 40 atoms and simultanously displacing one by one atom about 0.06 Å from its orginal position. To calculate the second-order harmonic and third-order anharmonic force constants (IFCs), the atomic forces for each atomic displacement were calculated by using plane wave pseudopotential method and PBE-GGA [26,27] functional in Quantum espresso [28]. For this, the convergence criteria were set to $10^{-10}$ and the criterion of force convergence was set to $10^{-4} eV/Å$. The kinetic energy cutoff 372 eV for wavefunctions and 1088 eV for charge density were used. In the IFCs calculation, ultrasoft pseudopotential (generated by PS library 1.00) was utilized. Finally, lattice thermal conductivity was calculated by performing BZ integration in the q-space with $21 \times 21 \times 21$ mesh and the equation $\kappa = \frac{1}{NV} \sum_\lambda C_\lambda \boldsymbol{v}_\lambda \otimes \boldsymbol{v}_\lambda \tau$, where V is the volume of the unit cell, $\boldsymbol{v}$ is the group velocity, $\tau$ is the SMRT for the phonon mode λ, and $C_\lambda$ is mode dependent phonon heat capacity. The relaxation time (τ) was calculated from the self-energy (Γ(ω)) as found in the above calculation by using $\tau = \frac{1}{2\Gamma(\omega)}$. This method has successfully predicted the lattice thermal conductivity of many compounds [29–32]. Since PBE-GGA underestimates the lattice thermal condcuctivity [30], the lattice thermal conductivity was also calculated by using GW hybrid functional [33–37].

## 3. Result and Discussions

### 3.1. Structural optimization

Using the experimental lattice parameters 5.1394 Å [18], we have created $2 \times 2 \times 2$ supercell containg total 40 atoms as illustrated in the Fig.1 (b).

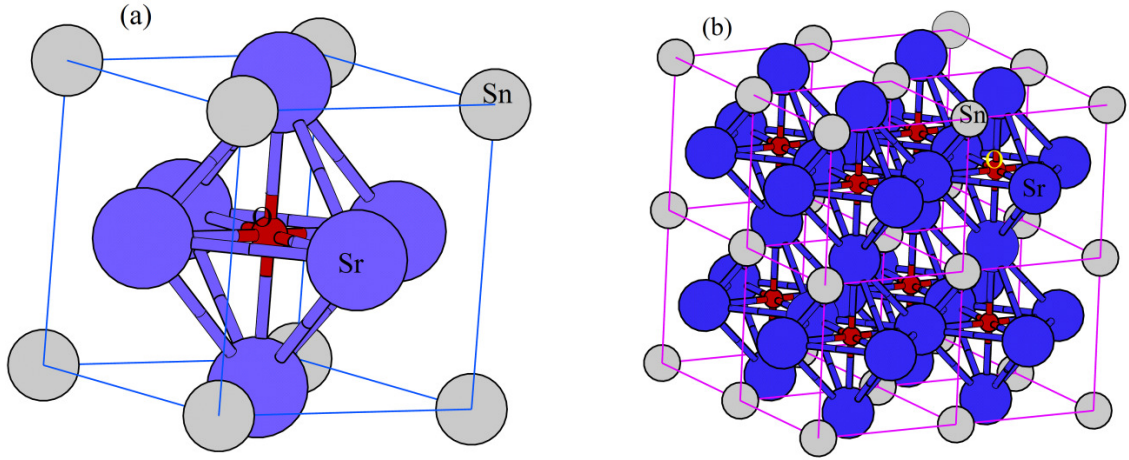

Fig.1: Crystal structure of (a) $Sr_3SnO$ (SSO) and (b) $2 \times 2 \times 2$ supercell structure of $Sr_3SnO$. These figures are sketched by using XCrySDen program[38].

After this, we have replaced one Sr atom with Ca (x=0.125) and applied the symmetry operation. The symmetry operation has reduced the nonequivalent atoms to 13 and lowered the symmetry to the space group $P4/mmm$ (#123). We have performed geometry optimization to find the equilibrium configuration. By applying the same procedure, for $x=0.25$ the space group has been found to be $P4/mmm$, for $x=$ 0.375 0.5, 0.62, 1,2 has been found to be $Pmmm$ (#47), and for $x=3$, the space group has been found to be $Pm\bar{3}m$ (#221). The optimized lattice parameters for $x=3$ ($Sr_0Ca_3SnO$, i.e., $Ca_3SnO$) has been found to be 4.8373 Å with excellent agreement with the experimental value 4.827Å [18]. Since the PBE-GGA is sufficient for structural optimization, the geometry optimization was performed with PBE-GGA functional. By using the equilibrium configuration, we have performed all other calculations (with TB-mBJ functional) as mentioned in the next sections.

## 3.2. Electronic structure

The calculated energy band structures of $Sr_{3-x}Ca_xSnO$ are illustrated in the Fig. 2. For $x=0$, our calculated band structure is consistent with others calculations[1,4,39]. However, the Fermi level is slightly (less than 0.05 eV) lowered than that obtained by including spin-orbit coupling (SOC) the effectin the calculation [4,39].Our calculated bandgap of $Sr_3SnO$ ($x=0$) by using TB-mBJ is consistent with the reference [11]. Another fact that stoichiometric $Sr_3SnO$ has been experimentally found to exhibit semiconducting resistivity[1,5,7]. The "Dirac cone" exists along the Γ-X direction with direct bandgap 0.22 eV as shown in the Fig. 2(a). When one Sr is repleced by Ca ($x=0.125$), the Fermi level significantly shift to the higher energy and increase the parabolic nature of the conduction bands with increasing bandgap as shown in the Fig. 2(a).

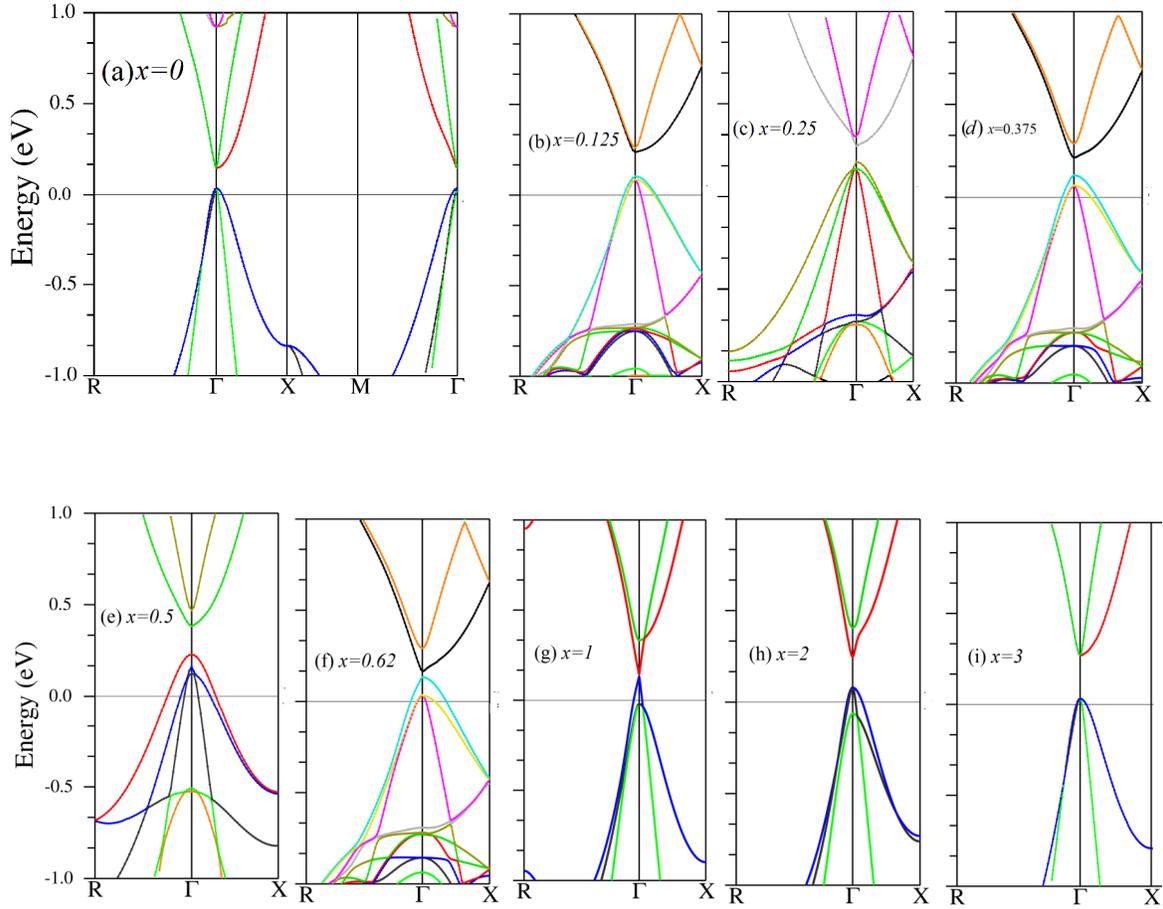

Fig. 2: Energy band structure of: (a) $Sr_3SnO$ and (b-i) its alloys $Sr_{3-x}Ca_xSnO$. The solid line at the zero energy represents the Fermi level. For $x$=0.125, 0.25, 0.375, 0.5, 0.62, the band structures are plotted for $2 \times 2 \times 2$ supercell.

This behavior is consistent with other semiconducting materials [40–44]. To take account of this effect further, we have calculated the effective mass of conduction band minima and valence band maxima by takeing second E-K derivative using finite difference method. Details of calculation method can be found in the reference [45]. The results are listed in the Table-1. We see that due to the Ca substitution ($x=0.125$), the holes effective mass increase with the bandgap. By increasing Ca-concentration in SSO, the conduction band (CB) and valence band (VB) becomes more parabolic and maximum parabolicity obatined for $x=0.5$.

Table-1: The calculated energy bandgap, effective mass of electrons, and holes in the unit of $m_0$, where $m_o$ is the bare electrons mass for different concenrations of Ca in SSO.

|  |  | x=0 | x=0.125 | x=0.25 | x=0.375 | x=0.5 | x=0.625 | x=1 | x=2 | x=3 |
|---|---|---|---|---|---|---|---|---|---|---|
| Eg (eV) | | 0.22 | 0.37 | 0.49 | 0.35 | 0.62 | 0.33 | 0.30 | 0.37 | 0.32 |
| m* | Electron | 0.14 | 0.694 | 0.629 | 0.47 | 0.75 | 0.25 | 0.0101 | 0.2 | 0.09 |
| | Hole | 1.42 | 1.638 | 1.802 | 1.11 | 0.88 | 1.37 | 2.01 | 1.55 | 1.78 |

At this concentration, we see that the effective mass of electrons significantly increases by reducing effective mass of holes. For $x=0.375$ and $0.625$, the bandgap decreases compairatively to $x=0.5, 0.25, 0.125$ as well as the effective mass of the electrons. The maximum hole effective mass is obatined for $x=1$. The hole effective mass of CSO (1.78) is reaonably close to the value (1.96) obatined from the experimental data [10]. The density of states (DOS) of SSO and its alloys are illustrated in the Fig. 3. The density of states is larger for $x=0.125, 0.375, 0.625$ than other concentrations. However, the DOS for $x=0.5$ is also compareable to DOS for these concentrations at the Fermi level. We see that main features of the density of states after Ca substitution remains almost same and only Fermi level shifts which is consistent with our band structure calculation. The calculated projected density of states of SSO and CSO is shown in the Fig. 4. The peak around -1 eV arises from the strong hybridyzation between Sr-4d/Ca-3d

(unoccupied states) and Sn-5p. The corresponding peak at the positive energy comes from the antibonding combinations. The oxygen has no contribution to the density of states around the Fermi level.

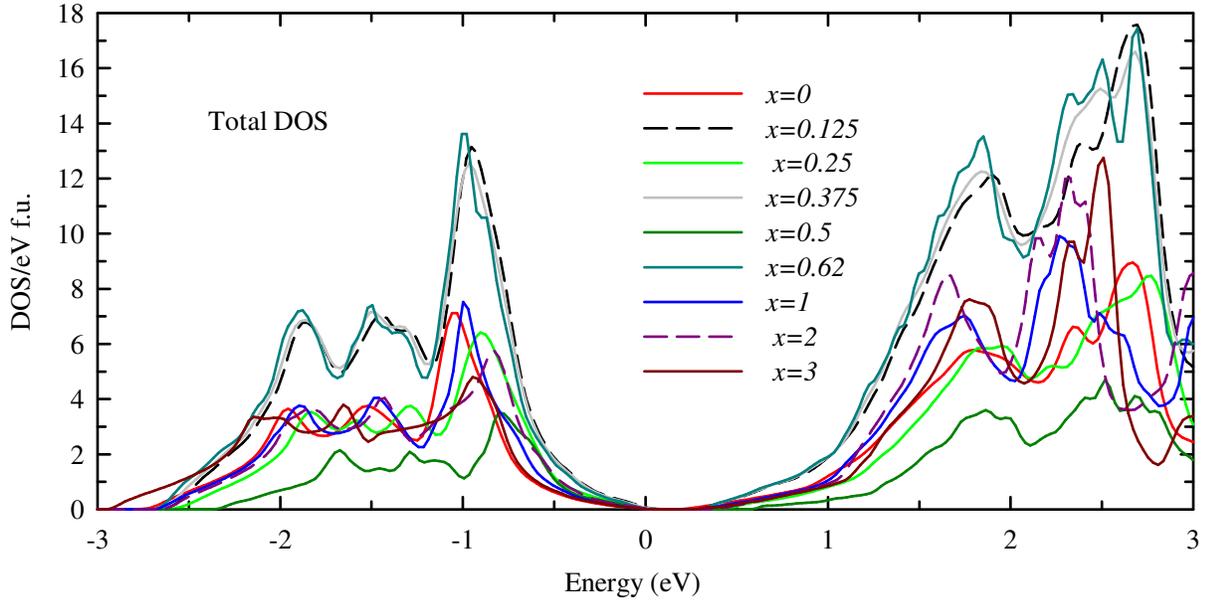

Fig. 3: Total density of states (DOS) of $Sr_{3-x}Ca_xSnO$. The Fermi level is set to zero energy.

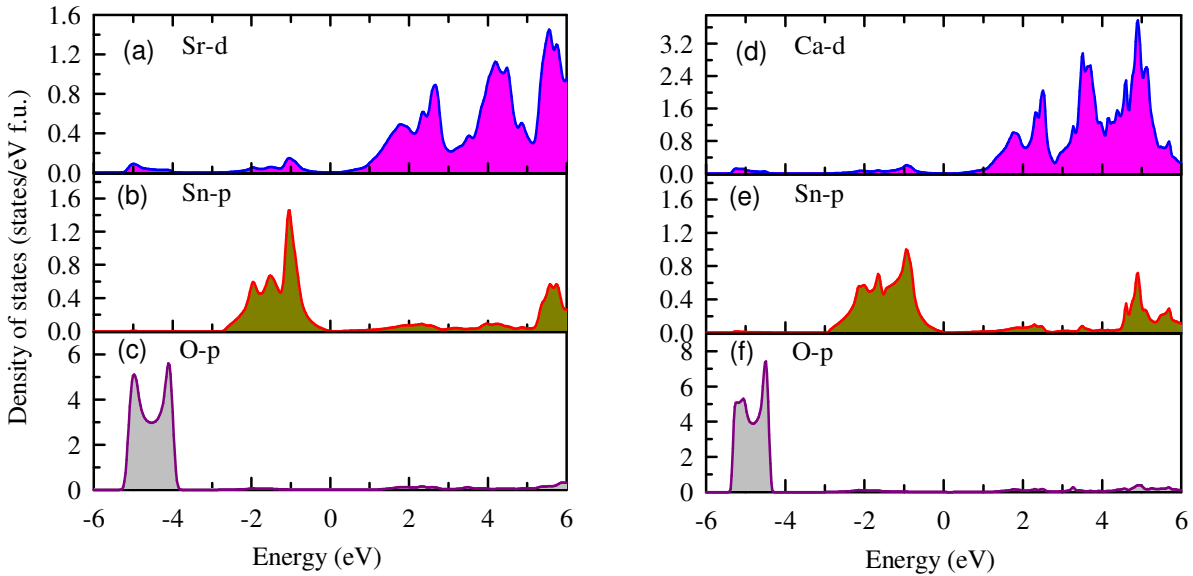

Fig.4: Projected density of states of $Sr_3SnO$ (x=0) and $Ca_3SnO$ (x=3). The zero energy represents the Fermi level.

## 3.3. Phonon transport properties

The phonon relaxation time of $Sr_3SnO$ and $Ca_3SnO$ decreases with increasing temperature due to the increase of phonon scattering as shown in the Fig. 5(a). At low temperature, relaxation time is large as scattering is significantly low. The theoretical Debye temperature ($\Theta_D$) of $Sr_3SnO$ is 276.1K [2], therefore, the scattering of electrons above $\Theta_D$ should be almost entirely due to phonons as acoustic phonon scattering is dominant in this range of temperatures [46,47]. In this range of temperature, the relaxation time shows $T^{-3/2}$ dependence[48] as shown in the inset figure of (a). Such linear dependence (inset figure) implies that the acoustic phonon scattering is predominant process in these antiperovskite. González Romero*et al* sugessted that $T^{-3/2}$ dependence may be used to find the relaxation time at different temperature[49].

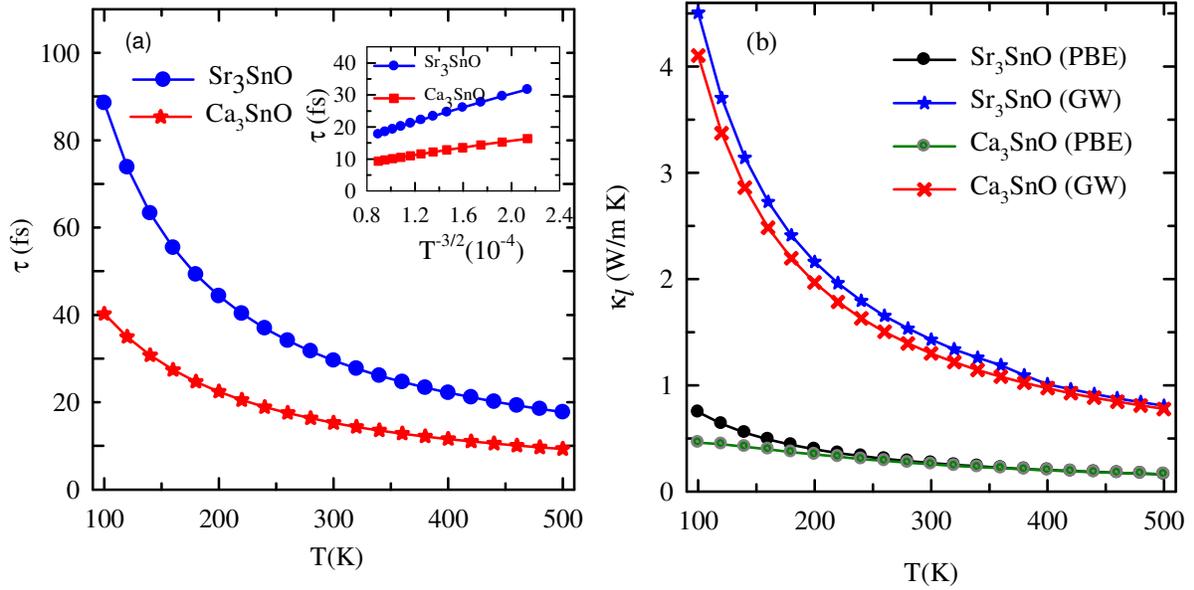

Fig. 5: The calculated: (a) relaxation time from phonon self-energy by using PBE functional, and (b) lattice thermal conductivity by using both PBE and GW functionals. The relaxation time obtained by GW is almost same and thus it is not shown in the Fig. 5(a) for clarity.

Our calculated relaxation time for SSO and Ca$_3$SnO (CSO) is in a reasonable range for typical thermoelectric materials and exhibits similar the temperature dependence[50,51]. The calculated lattice thermal conductivity (κ$_l$) of SSO and CSO as shown in the Fig. 5 (b). The calculated lattice thermal conductivity by using PBE is small for both compounds the value is 0.268 and 0.257 W/m K at 300 K for SSO and CSO, respectively. The experimental total thermal conductivity of CSO at 1.7 W/m K at 290K in which the lattice contribution is predominant[10]. Therefore, PBE method underestimates the lattice thermal conductivity of CSO. To overcome this underestimation, we have calculated it by using GW method as shown in the Fig. 5(b). and the κ$_l$ for both compounds is much larger than that obtained by PBE. The value of κ$_l$ is 1.43 and 1.3 W/mK for SSO and CSO, which is about 5 times larger than by PBE. Thus, the lattice thermal conductivity of CSO is consistent with the observation that the contributions of phonons to the thermal conductivity is predominant up to 380 K, as illustrated in the Fig. 9 (b). The Ca substitution should induce lattice distortions by shortening the mean free path of phonons. Thus, lattice thermal conductivity should be decreased by Ca substitution [52,53], and hence we have not performed the calculation for other Ca substituted SSO alloys.

## 3.4. Electron transport properties

The electron transport properties of a material mainly depend on its carrier concentration. For good thermoelectric materials, optimum carriers concentration is required [54–56]. The change of carriers concentration with Ca concentration (*x*) is illustrated in Fig. 6(a). The maximum carrier concentration ($3.471 \times 10^{19} cm^{-3}$) is obtained for x=0.5 and the minimum ($0.859 \times 10^{19} cm^{-3}$) is obtained for *x*=1. The carrier concentration for *x*=3 (in CSO) ($1.37 \times 10^{19} cm^{-3}$) is in the reasonable order with the experimental value ($1.43 \times 10^{19} cm^{-3}$)[10]. However, CSO exhibit metallic behavior due to Ca-deficiency and we will clear it on later in this section. Therefore, experimental carrier concentration should be larger than our semiconducting pure single crystal CSO. The variations of Seebeck coefficient with Ca concentration (x) are illustrated in Fig. 6(b).

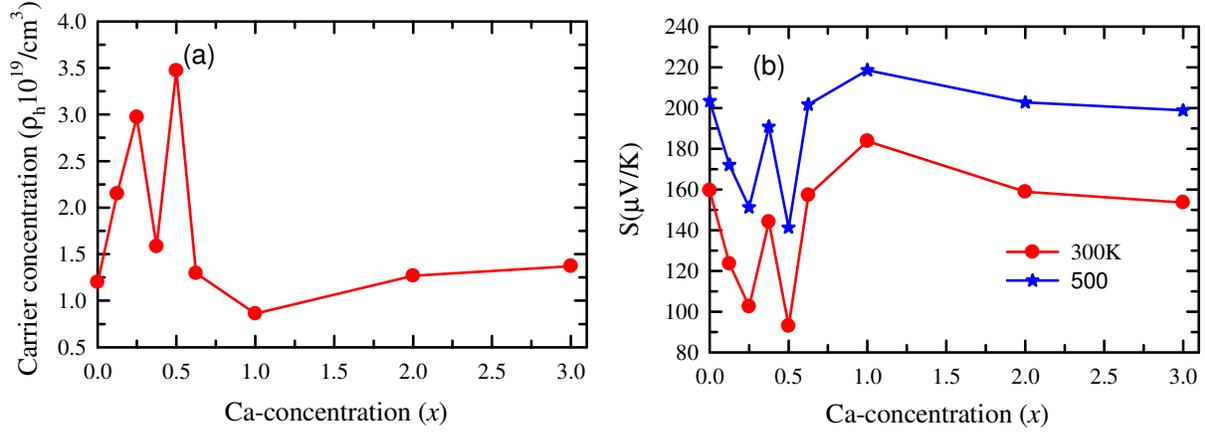

Fig.6: (a) Change in carrier concentration with Ca concentration ($x$), and (b) Seebeck coefficient with doping concentration at 300 K and 500 K.

The maximum Seebeck coefficient (S) is obtained for $x=1$ for which the carriers concentration is minimum. This indicates that the thermoelectric performance would be significantly increased by reducing the carrier concentration. Due to the increase of carriers concentration, the Seebeck coefficient is significantly reduced at x=0.5. The positive Seebeck coefficient (S) implies that the holes are the predominant carriers for these compounds. This is consistence with the experimental observation for CSO [10] and also with our calculated effective mass (m*). Such high effective mass with low carriers concentration is the normal belief to lead high Seebeck coefficient [57–61]. The maximum Seebeck coefficient is obtained for maximum effective mass at the minimum carriers concentration at $x=1$. High effective mass implies the low mobility of the carrier and high Seebeck coefficient although power factor may be reduced [62]. The calculated transport parameters of Ca substituted SSO alloys at different temperature are illustrated in Fig. 7. The Seebeck coefficient of p-type material can be expressed as $S = \frac{E_C - E_F}{q_h} + \frac{2k_B}{q_h}$, where $E_C$, $E_F$, $k_B$, and $q_h$ are the minimum energy of the conduction band, Fermi level, Boltzmann constant, and the hole charge, respectively, at the absolute temperature T. Due to the increase of electron concentration in semiconductors, the Fermi level generally shifts toward the conduction band which can be clearly understood from the band structure as shown in the Fig. 2. For such shift of Fermi level, the energy difference between the minima of the conduction band and Fermi level ($E_C$-$E_F$) becomes smaller and thus the equation (1) says that the minimum Seebeck coefficient will result. When the temperature is increased, the Fermi level shifts toward the middle of the bandgap and thus energy difference between them increases resulting in the increase of the absolute Seebeck coefficient. Thus, the Seebeck coefficient for SSO, CSO and their alloys increases with increasing temperature as shown in Fig. 7(a). However, the Seebeck coefficient of SSO increases up to 350 K and then decreases due to its activation energy. Moreover, for a certain temperature Fermi level is constant and after this temperature, it decreases. The obtained maximum Seebeck coefficient is 219 µV/K at 500 K with Ca concentration *x*=1. The reason behind the study of transport properties up to 500 K is that SSO changes its phase from cubic to orthorhombic[18]. The electrical conductivity of SSO and its alloys increases with increasing temperature implying the semiconducting nature of them. The electrical conductivity for x=0.25 increases very slowly with temperature due to its low carrier concentration (see Fig. 6(a).). By using the relaxation time ($2.956 \times 10^{-14}$ s) at 280 K for SSO, the electrical conductivity is $5.02 \times 10^4$ S/m and the resistivity is $1.990 \times 10^{-5}$ Ωm, with the reasonable agreement with the experimental value $6.750 \times 10^{-4}$ Ωm, measured on polycrystalline chunks.

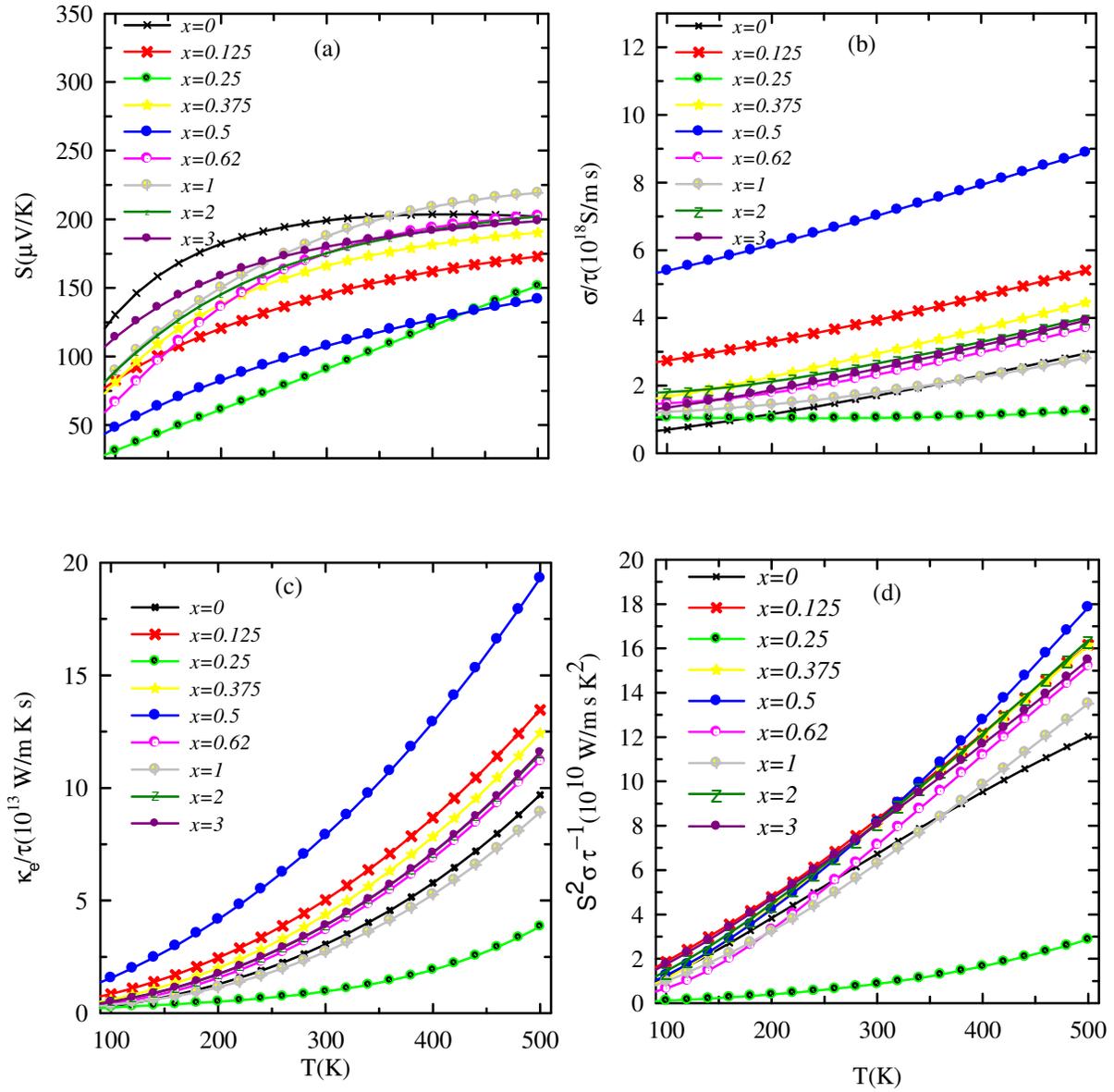

Fig. 7: Electronic Transport properties of $Sr_{3-x}Ca_xSnO$: (a) Seebeck coefficient, (b) electrical conductivity ($\sigma/\tau$), (c) electronic part of the thermal conductivity ($\kappa/\tau$), and (d) power factor ($S^2\sigma/\tau$).

The electronic part of the thermal conductivity increases with temperature as usual due to the increase of electrical conductivity as shown in the Fig. 7(c). The calculated power factor (PF) at different temperature are illustrated in the Fig. 7(d). The PF increases with increasing temperature as expected. Although maximum power factor is obtained for $x=0.5$, the thermoelectric figure of merit ZT should small due to high electronic thermal conductivity. The question may arise for CSO that our calculate electrical conductivity show semiconducting behavior while the experimental sample of the reference [10] shows metallic nature. To find out the reason behind this, we have studied Ca, Sn, and also O-deficiency. Due to the Sn-deficiency, the CSO becomes an n-type material which is not consistent with the experimental observation. Due to the O-vacancy, the material exhibit metallic conductivity but very small Seebeck coefficient. However, the Ca-deficiency give rise to the metallic conductivity also large Seebeck compared to O-vacancy. The calculated Seebeck coefficient and electrical conductivity of CSO with Ca-deficiency (12.5%, i.e., $Ca_{2.875}SnO$) are shown in the Fig. 8. Our calculated S and $\sigma/\tau$ show a similar trend as experimentally observed for CSO [10].

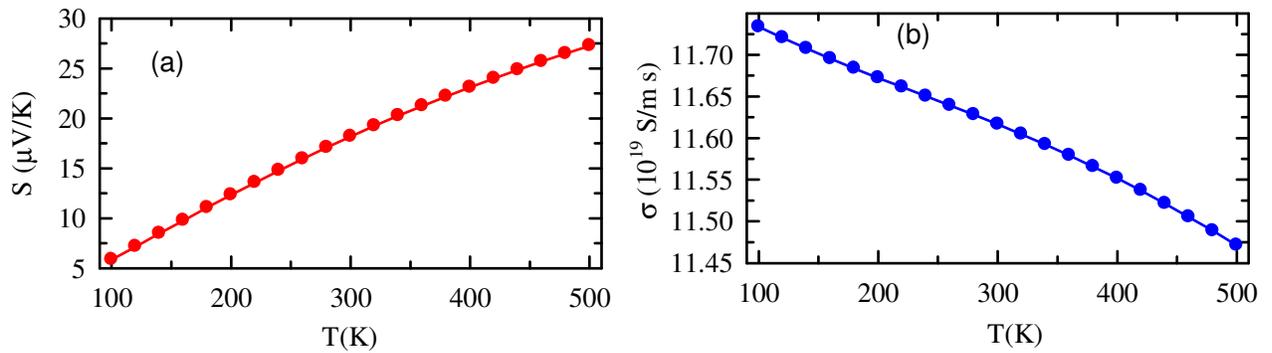

Fig. 8: (a) Seebeck coefficient (S) and (b) electrical conductivity of $Ca_3SnO$ with Ca-deficiency (12.5%, i.e., $Ca_{2.875}SnO$).

The electrical conductivity is $1.899 \times 10^6$ S/m by using the relaxation time at 280K ($1.6334 \times 10^{-14}$ s), and the resistivity is $5.265 \times 10^{-7}$ Ωm, with the reasonable agreement with the experimental value $6.348 \times 10^{-5}$ Ωm, measured on polycrystalline sample. Therefore, it is important to synthesis pure CSO to obtain high thermoelectric performance. The calculated total thermal conductivity of SSO and CSO are presented in the Fig. 9. The total thermal conductivity decreases with increasing temperature as the lattice contribution is large. However, the total thermal conductivity obtained by using the PBE calculated lattice thermal conductivity is much smaller than that with GW method. The total thermal conductivity (κ) of CSO is reasonably close to the experimental value. Our calculated value is slightly larger because the experimental measurement was performed on the polycrystalline sample in which phonon scattering is small due to grain boundary than the single crystal.

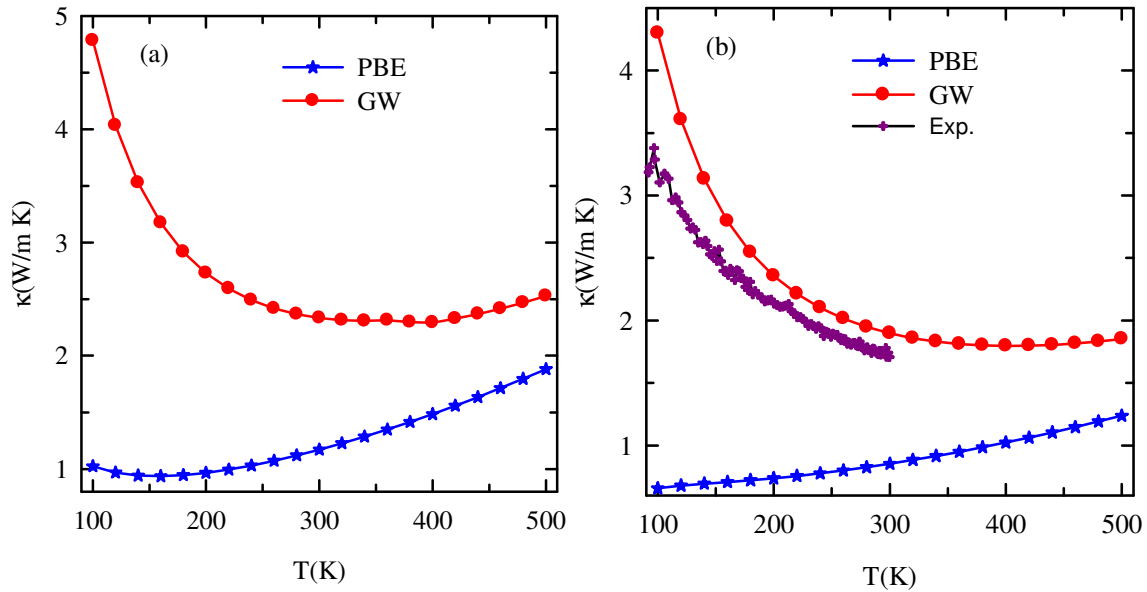

Fig. 9: Total thermal conductivity of (a) $Sr_3SnO$ and (b) $Ca_3SnO$ by using the lattice thermal conductivity obtained through PBE and GW methods. The experimental data are taken from the reference [10].

The κ shows $T^{-1}$ dependence for both compounds as experimentally observed for CSO [10]. The κ of both compounds is relatively larger than that for glasses and amorphous alloys (0.5-1 W/mK) [63,64] but relatively smaller for inorganic compounds [65]. Low thermal conductivity is a very crucial factor for the high thermoelectric figure of merit (ZT).

### 3.5. Figure of merit

The dimensionless thermoelectric figure of merit (ZT) at different temperature of SSO and its alloys are presented in Fig. 10. The ZT value increases with increasing temperature and the maximum value for SSO is 0.42 at 500 K. As the Ca concentration, x increases from $x=0.125$, ZT decreases. But surprisingly increases for $x=0.375$ due to low carrier concentration (high Seebeck coefficient). The minimum ZT is obtained for $x=0.25$ due to high carrier concentration, and large bandgap (reducing the electrical conductivity).

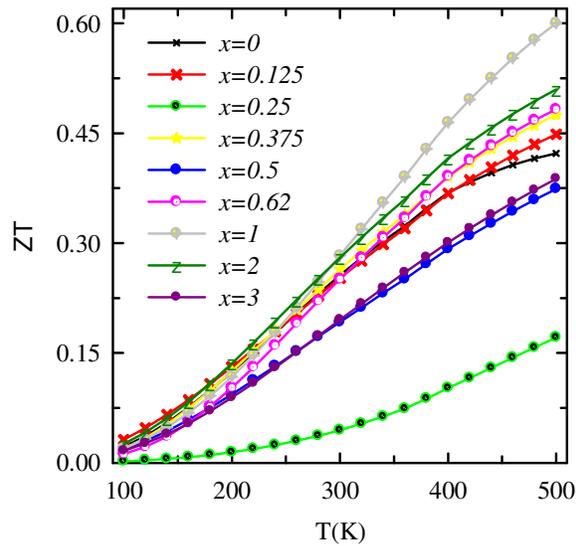

Fig. 10: Variations of the dimensionless thermoelectric figure of merit of $Sr_{3-x}Ca_xSnO$ with temperature.

The maximum ZT (0.6) is obtained for *x*=1, lowering the carrier concentration, due to high effective mass (high Seebeck coefficient) and optimum bandgap. The ZT for *x*=1 (0.6) is 1.45 times larger than the pure SSO crystal. The experimentalists should be interested to study on this further to find the optimum carrier concentration and high thermoelectric performance with remembering that the pure SSO crystal is the essential criteria for high ZT in these compounds.

## 4. Conclusions

In summery, we present a comprehensive set of the first-principles study of the electronic and thermoelectric transport properties of Ca substituted $Sr_3SnO$. We have found that the Fermi level shifts toward the conduction band due to the Ca substitution. The maximum bandgap has been obatined for *x*=0.5 (0.62 eV) for which the effective mass of electrons is maximum (0.75). The minimum holes effective mass (1.11) has been obatined for *x*=0.375. We have found that the thermopower (Seebeck coefficient) of SSO and also the effective mass of holes are significantly reduced due to the increase of hole concentration. The optimum hole concentration has been obtained for *x*=1 and the corresponding effective mass of holes and thermoelectric performance is maximum. The maximum Seebeck coefficient is 219 µV/K. The electrical conductivity of SSO and its alloys exhibits semiconducting nature which contradicts with experimental observation for $Ca_3SnO$. To overcome this, we have studied the Seebeck coefficient and electrical conductivity of defect CSO. We have found that due to the Ca-deficiency, the CSO shows the metallic conductivity and removes this contradiction with our theoretical results. This reveals the importance of synthesis of pure CSO to obtain high thermoelectric performance. The lattice thermal conductivity ($\kappa_l$) of SSO and CSO has been calculated by using both PBE and GW functionals. The lattice thermal conductivity obtained by PBE functional largely underestimates the experimental value for CSO. The total thermal conductivity (with $\kappa_l$ obtained by GW) at 300K is 2.33 and 1.897 W/m K for SSO and CSO, respectively, with excellent agreement with experimental value 1.707 W/m K for CSO. The maximum thermoelectric figure of merit (ZT) is 0.6 at 500 K for *x*=1, which is 1.45 times larger than that for the pristine SSO (0.42 at 500 K). We conclude that the thermoelectric performance in Ca substituted $Sr_3SnO$ could be further improved by nanostructuring and grain refinement to reduce the lattice thermal conductivity without affecting the electrical conductivity. We hope that our obtained results will inspire further experimental study on the thermoelectric properties in Ca substituted $Sr_3SnO$.